\documentstyle[referee]{mn}

\newcommand{\ha}{H$\alpha$}

\newcommand{\degree}{$^{o}$~}
\def\vhel{\ifmmode{V_{{\rm HEL}}}\else{$V_{{\rm HEL}}$}\fi}
\def\vsys{\ifmmode{V_{\rm sys}}\else{$V_{\rm sys}$}\fi}
\def\kms{\ifmmode{~{\rm km\,s}^{-1}}\else{~km~s$^{-1}$}\fi}
\def\vlsr{\ifmmode{v_{\rm lsr}}\else{$v_{\rm lsr}$}\fi}
\title[Eridanus shells] {Deep \ha\ imagery of the Eridanus shells}

\author[P. Boumis et al] {P. Boumis$^{1,2,3}$, C. Dickinson$^{1}$,
J. Meaburn$^{1}$, C.D. Goudis$^{4}$, P.E. Christopoulou$^{4}$, \cr
J.A. L\'{o}pez$^{5}$, M. Bryce$^{1}$ \& M. P. Redman$^{1}$ .\\
$^{1}$Jodrell Bank Observatory, Dept of Physics \& Astronomy,
University of Manchester, Macclesfield, Cheshire SK11 9DL UK.\\
$^{2}$Department of Physics, University of Crete, P.O. Box 2208,
GR-710 03 Heraklion, Crete, Greece.\\$^{3}$Foundation for Research and
Technology-Hellas, P.O. Box 1527, GR-711 10 Heraklion, Crete,
Greece.\\ $^{4}$Astronomical Laboratory, Department of Physics,
University of Patras, 26500 Rio-Patras, Greece.\\ $^{5}$ Instituto de
Astronom\'{\i}a, UNAM, Apdo. Postal 877, Ensenada, B.C. 22800,
M\'{e}xico.\\ }

\date{Received
**insert**; Accepted **insert**}
\pagerange{\pageref{firstpage}--\pageref{lastpage}}

\begin{document}  

\maketitle

\begin{abstract}
 
\noindent A deep \ha\ image of interlocking filamentary arcs of
nebulosity has been obtained with a wide-field ($\approx$ 30\degree
diameter) narrow-band filter camera combined with a CCD as a detector.
The resultant mosaic of images, extending to a galactic latitude of
65$^{o}$, has been corrected for field distortions and had galactic
coordinates superimposed on it to permit accurate correlations with
the most recent H{\sc i} (21 cm), X-ray (0.75 kev) and FIR (IRAS 100 $\mu$m)
maps.

Furthermore, an upper limit of 0.13 arcsec/yr to the expansion proper
motion of the primary 25\degree\ long nebulous arc has been obtained
by comparing a recent \ha\ image obtained with the San Pedro Martir
telescope of its filamentary edge with that on a POSS E plate obtained
in 1951.

It is concluded that these filamentary arcs are the superimposed
images of separate shells (driven by supernova explosions and/or
stellar winds) rather than the edges of a single `superbubble'
stretching from Barnard's Arc (and the Orion Nebula) to these high
galactic latitudes. The proper motion measurement argues against the
primary \ha\ emitting arc being associated with the giant radio loop
(Loop 2) except in extraordinary circumstances.

\end{abstract}


\section{Introduction}

A 25\degree diameter area of interlocking arcs of \ha\ emitting
filamentary nebulosity was discovered (Meaburn 1965, 1967) projecting
from the galactic plane to high galactic latitudes ($b = -50^{o}$). The
primary nebulous arc (hereafter called Arc A) is traced from $(l,b) =
(180^{o}, -30^{o})$ to $(l,b) = (200^{o},-50^{o})$ (in the Eridanus
constellation) and has a similar morphology to the emission regions
associated with old supernova remnants (Meaburn 1967). For this reason
an association of Arc A with one of the giant radio loops (the
90\degree diameter Loop 2, Large et al. 1962, Haslam et al. 1971) was
suggested, in which case, Arc A could have been within 30 pc of the
Sun. Reynolds \& Ogden (1979), henceforth RO, though suggested that
the arcs were edges of a single elongated cavity expanding at
$\approx$ 15 \kms\ and extending from the I-Ori OB-association and
therefore at the greater distance of $\approx$ 450 pc.  They cite the
proximity of the well-known `Barnard's Arc' (see Sivan 1974) in
support of this contention. Since this early work the Eridanus
nebulosity has continued to receive considerable attention in many
other wavelength domains largely because an uncluttered view of the
closest galactic interstellar shells is permitted at such high
galactic latitudes.

More recently, data from the Leiden/Dwingeloo H{\sc i} survey (Hartmann \&
Burton 1997, Burton \& Hartmann 1994), the IRAS Sky Survey Atlas ISSA
(Wheelcock et al. 1994) and ROSAT X-ray survey (Burrows \& Guo 1996,
Snowden et al. 1995) have provided a more detailed picture of the
region. Comprehensive comparisons between the data sets have been made
by Burrows et al (1993), Brown et al (1995), Heiles et al (1998) and
Heiles et al (1999). They suggest a similar unifying picture of the
region similar to that given by RO - a large superbubble (SB) filled
with hot X-ray emitting ($T \approx 1 \times 10^{6} K$) gas extending
from the Ori OB1 association at a distance of $\approx 450$pc, with
the nearest edge at $\approx 150$pc. Burrows \& Guo (1996) estimated a
distance of $159\pm16$pc in the direction
$(l,b)=(200^{o},-47^{o})$. The cavity is enclosed in an expanding
neutral shell traced between $-40 \kms \ge \vlsr \ge +40 \kms$ which
is seen preferentially edge-on. However, the structure of the
`Eridanus-Bubble' and its formation are still under scrutiny. Heiles
et al (1999) has pointed out that it is clear that there are some
dependencies on other nearby shells including the local
`superbubble' (SB), in which the Sun is immersed.  However, the
filamentary Arc A does not fit clearly into this simple SB
picture. For one thing, its curvature is in the opposite sense to that
predicted by the SB model.

Deep, narrow-band, \ha\ images have been obtained with a wide-field
CCD camera of the Eridanus nebulosity to permit a more detailed, and
astrometrically accurate, comparison with features at other
wavelengths. Furthermore, a new upper limit to the proper motion (pm)
of the sharp filamentary edge of Arc A has been obtained in an attempt
to constrain the lower limit to its distance.  The previous
measurement (Jones \& Meaburn 1985) used material over a shorter
baseline and with a lower signal-to-noise ratio.

\section{Observations and Results}

\subsection{Wide field imagery}

The layout of the Manchester Wide Field camera (MWFC) is shown in
Fig. 1. A 21\degree $\times$ 30\degree field-of-view is converted to
5.5\degree $\times$ 7.8\degree for acceptance in the parallel beam by
a three-period (square profile) interference filter, of 16\AA\
bandwidth centred on \ha. The field is finally imaged on to a Wright
Instrument, nitrogen cooled, CCD with 385 $\times$ 578 pixels giving a
scale of $\sim$ 3.29\arcmin ~pixel$^{-1}$. The observations reported
here were made at Kryonerion Observatory in Greece, between 12--16 of
December 1996.

The eight fields covered and the integration times used are listed in
Table~1.  The negative grey-scale representations of the final mosaic
is shown in Fig.~2. The galactic coordinates are accurate to a few arcmins
for the images were processed (Boumis 1997) using the Starlink {\sc kappa}
packages {\sc centroid} and {\sc setsky}, the {\sc iras90} packages
{\sc skyalign} and {\sc skygrid}
and the mosaicing {\sc ccdpack} package {\sc makemos}. All the astrometry
information can then be displayed in the final image using {\sc skygrid}. In
particular, optical field distortions were corrected in this process.
 
\begin{table*}
\centering
\begin{tabular}{||c|c|c|c|c||}
\hline
Position &
Date &
Exposure Time &
R.A. &
Dec. \\
N$^{o}$ &  & (min) & (h m) & ($^\circ$) \\
\hline
1 & 13 December 1996 & 20, 30 & 3 00 & 0 \\ 
\hline
2 & 12 December 1996 & 20, 30 & 3 30 & 0 \\
\hline
3 & 16 December 1996 & 30 & 3 40 & -5 \\
\hline 
4 & 13 December 1996 & 30, 30, 30 & 3 40 & 0 \\
\hline
5 & 12 December 1996  & 20, 20 & 4 00 & -5 \\
\hline
6 & 12 December 1996 & 20, 20 & 4 00 & +5 \\
\hline
7 & 16 December 1996 & 30 & 4 20 & -5 \\
\hline
8 & 12 December 1996 & 20 & 4 30 & +15 \\
\hline
\end{tabular}
\caption{Observational details for the eight separate fields 
the  Eridanus nebulosity.}
\label{table1}
\end{table*}

The \ha\ emitting regions in Fig. 2 are compared with the 100 $\mu$m
IRAS emission in Fig. 3(a) and with both the ROSAT X-ray (kev band)
emission and the H{\sc i} 21 cm ring (Brown et al. 1995) in Fig. 3(b). The
principal \ha\ arcs are called Arcs A \& B in Fig. 3(a).

\subsection{Proper motions}

The brightest part of the filamentary Arc A has a sharp edge in the
light of \ha. This appears somewhat faintly on the Palomar Observatory
Sky Survey (POSS) Schmidt plate E232, taken on the 3rd of January
1951.  Nonetheless the area containing Arc A was digitized with the
PDS scanning microdensitometer at the Royal Greenwich Observatory
(RGO). The resulting image measured 250 $\times$ 250 pixels with a
resolution of 20 $\mu$m ($\equiv$ 1.34\arcsec).

In an attempt to measure any proper motion of Arc A this same
filamentary edge was imaged (see Fig. 4) on the 8th of November 1996
with the 2.1m San Pedro Martir (SPM) telescope in Mexico. The
integration time was 1800s through an interference filter centred \ha\
and with an 11\AA\ bandwidth with the Tek CCD, (1024 $\times$ 1024, 24
$\times$ 24 $\mu$m$^{2}$ pixels to give a scale of
0.3\arcsec~pixel$^{-1}$ at the f/8 focus) as the detector. The
resultant `seeing' disk was 1.8\arcsec\ diameter nearly matching the
POSS resolution.

The Starlink {\sc ccdpacl} package {\sc ccdalign} was used to align the two data
arrays and the {\sc gaia} package {\sc patch} to then remove the star images. Both
arrays were finally rotated by the same amount until the filamentary
edge was vertical in both.  Brightness profiles were compared for a
cut, 2.15\arcmin\ long and marked in Fig. 4, perpendicular to the most
defined part of the filamentary edge of the nebulous Arc A.  These are
shown in inserts in Fig. 4.  A conservative upper limit to the proper
motion, perpendicularly to the filamentary edge, of $\pm 0.13$\arcsec\
yr$^{-1}$ was given with the detection limit governed by the lower
signal-to-noise ratio in the POSS data.

\section{Discussion}

There is an immediate implication of the expansion pm upper limit for
Arc A in Sect. 2.2. Consider a pm of $\delta \theta$ arcsec yr$^{-1}$
for a filament with a tangential velocity of V$_{\rm t}$ \kms\ at a
distance D pc then D = 4.612 V$_{\rm t}$ $\delta \theta^{-1}$. If Arc A is
the edge of a shell expanding slowly at 15 \kms\ then for an expansion
pm of $\leq$ 0.13 arcsec yr$^{-1}$ its distance must be $\geq$ 530 pc.
The 25\degree long Arc A would have a linear extent of $\geq$ 230 pc
with this lower distance limit.  Any higher expansion velocities than
15 \kms\ , such as those of $\geq$ 50 \kms\ typical of old supernova
remnants (Cygnus Loop, IC443, Vela etc) and capable of causing
collisional ionization, would then place Arc A at the improbably high
distance of $\geq$ 1.8 kpc with a corresponding increase in linear
size.

The relationship of Arc B to both Barnard's Arc and Arc A should now
be considered.  Firstly, Heiles et al (1999) show that an area of soft
X-ray emission mapped by Snowden et al (1995) extends from Barnard's
Arc (15\degree diameter centred on $(\ell,b)= (224.5^{o}, -18^{o})$
and the Orion Nebula) to Arc B. In fact they describe this volume as
the `Eridanus superbubble' with Arc B as the supershell wall. The
detailed correlations shown in Fig.~3(b) suggest an alternative
interpretation.  Here, Arc B appears to be one edge of a
self--contained shell delineated by the 20\degree diameter ring of H{\sc i}
21cm emission surrounding a localised enhancement of super-heated
X-ray emitting gas.  Warm dust, giving rise to the 100 $\mu$m ridge
just outside Arc B, must be mixed with the neutral gas piled up as the
H{\sc i} shell expanded. A separate shell is suggested which appears to be
unrelated to Arc A and not to be part of a coherent, elongated
`supershell' emanating from the Orion region.  In fact this shell
could be one of a complex of shells whose overlapping images give rise
to the more extended X-ray region.

The mechanism for the ionization of the \ha\ emitting region
of Arc B must be considered within the model that this
ionized shell of gas is
on the {\it inside} surface of the larger 20\degree diameter
H{\sc i} shell. An expanding
supernova remnant, with a shock driven into the surrounding interstellar
medium would have exactly the reverse of this configuration
if its expansion velocity was great enough to generate a radiative
shock ie \ha\ emitting filaments would be expected
on its outside surface.
Furthermore, leakage Lyman photons from the Orion nebula are shielded
from this inside surface by the outer H{\sc i} shell. 
It can be seen in Fig. 3(b) that an arc of X-ray
emitting, super-heated, gas
is immediately adjacent to Arc B. This hot gas is most likely
the slowly cooling remnant of the shocked blast wave
that originally pressured the expansion of the shell
in its early, energy conserving, phase (assuming a supernova origin)
and is still present in the shell's present momentum conserving phase.

The X-ray emitting gas is most strongly located towards higher
galactic latitudes within the H{\sc i} shell. In general, the random
motions of H{\sc i} clouds do not allow them to rise to much more than
$100~{\rm pc}$ or so from the galactic plane. Hot ($\sim 10^6~{\rm
K}$) X-ray emitting gas will have a much larger scale height above the
galactic plane due to the high sound speed in gas of this temperature.
The scale height is about (Kahn~1998) ${c_{\rm H}^2}/g_{\rm z}$, where
$c_{\rm H}\simeq 100~{\rm km~s^{-1}}$ is the sound speed in the hot
gas and $g_{\rm z}\sim 10^{-8}~{\rm cm~s^{-1}}$ is the gravitational
field directed down towards the disk. This buoyancy of the hot gas
within the H{\sc i} shell may then have resulted in the hot gas gathering
at the top (with respect to the galactic plane) of the shell. If the
hot gas has enough pressure to eventually be able to break through the
H{\sc i} shell that is constraining it then it will be able to rise into the
galactic halo, forming part of the galactic fountain and leaving a
chimney structure behind (see e.\@g.\@~Norman~\& Ikeuchi 1989).
Photoionization
by the Lyman photons in the long wavelength wing of the $10^{7}$ K
Black Body emission from the super-heated gas is a  possible
ionizing mechanism. 
More likely the optical emission seen at the interface between the hot and cold
gas  arises as a result of mixing between the two
media. This can occur via thermal conduction (see e.\@g.\@~Borkowski,
Balbus \& Fristrom~1990, and references therein) as hot electrons heat
and ionize the inner edge of the H{\sc i} shell giving rise to a warm
layer of $\sim 10^5~{\rm K}$. This is a thermally unstable temperature
and the gas will rapidly cool to $10^4~{\rm K}$ resulting in the
emission of \ha. Thermal conduction
may be inefficient if significant magnetic fields are present that can
hinder the transport of hot electrons to the cold gas or if there is
significant relative motion between the two media (Hartquist \&
Dyson~1993). If this is the case, then mixing can take place via
localised boundary layers and shocks (Dyson~et~al~1993,
Redman~et~al~1999). An observational test that these effects are still
taking place would be the detection of O~{\sc vi} absorption lines
since this ion indicates temperatures of $3-5\times 10^5~{\rm K}$
which, as mentioned above, is a temperature from which material can
rapidly cool.

Incidentally, the best guide to understanding these structures extending from
Barnard's Arc to Arcs A and B in Fig. 2 and the extended X-ray emitting
gas must be the giant shells of the Large and Small Magellanic Clouds
and six interlocking shells between 68-308 pc diameter in the Local
Group galaxy IC~1613 (Meaburn et al. 1988).  For one thing their
dimensions correspond to those of Arcs A and B for distances $\geq$
530 pc.  The image of such a group of separate shells, viewed from the
edge of the group, but within the plane of a galaxy, would appear as
overlapping \ha\ emitting filaments along any sight-line or in X-rays,
by superimposed emission from separate volumes of super-heated gas. 

Arc A in Fig. 3(a \& b) could then be unrelated to Arc B and similarly
not part of an elongated `superbubble' projecting from Barnard's Arc:
it could simply be the manifestation of a further shell whose image is
superimposed on Arc B. The original correlation of Arc A with the
giant (90\degree diameter) non-thermal radio loop (Loop 2 - Meaburn
1965 and 1967) now seems unlikely in view of the present upper limit
to its expansion proper motion. Only if Loop 2 were a truly
fossilised, nearby, supernova remnant with an expansion velocity of
$\leq$ 10 \kms\ could Arc A exist within the requisite 30 pc of the
Sun. 
In this case ionization of the outer edge of a
compressed H{\sc i}, 90\degree diameter, swept--up
shell by leakage Lyman photons from the 450 pc distant Orion region
would also have to be invoked to produce the observed \ha\ emission.
The  sense of the curvature of Arc A is one feature in favour of this
interpretation. A relationship of Arc A with the nearby, separate H{\sc i}
shell found by Lindblad et al (1973) cannot be ruled out.

\section{Conclusions}

We calculate a lower limit of 530pc to the distance of Arc A by
assuming an expansion velocity of $\approx 15$\kms\ (RO, 1979). This
corresponds to a linear size of $\geq 230$pc. 
Arc B could be at a much
lower distance of $\approx 150$pc and seems to be  self-contained in a
20\degree diameter H{\sc i} ring. Therefore we have evidence for Arc A and
Arc B to be part of a complex of  individual shells viewed along the
same sight-line, and therefore not just the edge of a single Eridanus
superbubble. Further fainter arcs and loops can also be seen in the
deep \ha\ mosaic image at a lower level, supporting this
hypothesis. Furthermore, a complex of interlocking shells of this size
and morphology, is consistent with similar structures seen in the LMC
and SMC.

The mixing of extended soft X-ray emission with the H{\sc i} shell
wall (adjacent to Arc B) is argued to be the
most likely ionising mechanism for Arc B. 

Arc A on the
other-hand, could still be related to giant radio Loop 2, but
only for expansion velocities $\le10$\kms.

\section*{Acknowledgements} 

We wish to thank the staffs at the Kryonerion and
San Pedro Martir observatories  for their
excellent assistance during these observations and Dr.\ Bob Argyle of
RGO for assistance with the PDS microdensitometer. The work in Greece
was sponsored by the British Council (Athens) in a collaboration
between Patras and Manchester Universities. 
JAL acknowledges support from
DGAPA-UNAM and CONACY(Mexico) through projects IN114199 and 32214-E
respectively. CD acknowledges a PPARC research grant. MPR acknowledges a PPARC Research Associateship.

\section{Figure legends}

\noindent{\bf Figure 1. }

\noindent The optical layout of the Manchester wide field camera
(MWFC). The interference filter is in the exit pupil.

\noindent {\bf Figure 2.}
 
\noindent A negative grey-scale representation of the mosaic of \ha\ images
taken of the Eridanus nebulosity.

\noindent {\bf Figure 3(a).}
\noindent A comparison of the \ha\ contours (linear intervals)
from the image in Fig. 2 with a negative grey-scale representation
of the IRAS 100 $\mu$m image of the field. Arcs A \& B are indicated.

\noindent\noindent {\bf Figure 3(b).}
\noindent A lighter grey-scale representation of the \ha\ image in
Fig. 2 is compared with the ROSAT X-ray (0.75 kev) contours. The 
H{\sc i} ring is indicated by a dashed line from Brown et al (1995). 

\noindent {\bf Figure 4.}
 
\noindent A negative \ha\ image of the sharp edge of Arc B taken with
the San Pedro 2.1m telescope is shown.  This has been rotated to bring
the sharp edge of Arc A nearly vertical in the data array. The 
field centre is RA = 04h 01m 21.47s DEC = +01\degree 32\arcmin 20.5\arcsec   
(J2000) ($\ell ,b$) = (188.7\degree, --36.1\degree). Note that north
is to the top of the tilted left-hand edge of the image. Cuts taken
across the \ha\ filament from (a) the SPM telescope and (b) the POSS
data arrays are shown as inserts. These were co-additions from the respective
images for the length marked against the image of the filamentary
edge.
 
\bibliographystyle{mnras}

\end{document}